# Advection-dominated accretion: global transonic solutions

Xingming Chen[1], Marek A. Abramowicz[1], Jean-Pierre Lasota[2,3]

## ABSTRACT

We obtained global transonic solutions representing optically thin advection-dominated accretion flow by solving the full set of differential equations describing such systems. We found that far from the sonic point self-similar solutions are an excellent approximation of the global flow structure if the accretion rate is well below the maximum value above which no optically thin solutions exist.

*Subject headings:* accretion, accretion disks — black hole physics

## 1. INTRODUCTION

Accretion on to a compact object powers many astrophysical systems such as active galactic nuclei, X-ray binaries, and cataclysmic variables. It contributes to the energy sources of young stars (e.g., Bertout, Basri & Cabrit 1991) and plays a role in the physics of the nuclei of 'ordinary' galaxies (Fabian & Rees 1995) and of our Galactic Center (Narayan, Yi, & Mahadavan 1995). Until recently it was customary to assume, that the accretion power is radiated away with an efficiency of $\sim GM/(rc^2)$ (where $G$ is the gravitational constant, $c$ is the speed of light, $M$ is the mass of the central object, and $r$ the radial distance from it), which is $\sim 0.1$ close to the surface of a neutron star or a black hole. In other words it was assumed that most of the gravitational energy released through viscous dissipation say, was locally radiated away from the accretion disk. This condition is very well satisfied for a geometrically thin accretion disk (e.g., Frank et al. 1992).

Accretion flows in which radiative efficiency is very low and most of the heat gets advected into the central object have been studied by Begelman (1978, 1979) for spherical flows and by Begelman & Meier (1982), Abramowicz et al. (1986; 1988) for accretion disks. These disks were no longer geometrically thin (they were called 'slim disks' by Abramowicz et al. 1988) but were optically thick and corresponded to very high, super-Eddington accretion rates. The application of these models to real objects have not yet been found.

---

[1]Department of Astronomy and Astrophysics, Göteborg University and Chalmers University of Technology, 412 96 Göteborg, Sweden

[2]Belkin Visiting Professor, Department of Condensed Matter Physics, Weizmann Institute of Science, 76100 Rehovot, Israel

[3]UPR 176 du CNRS; DARC, Observatoire de Paris, Section de Meudon, 92195 Meudon, France

The interest in advection-dominated flows has been recently revived by the work of Narayan & Yi (1994, 1995a, 1995b), Abramowicz et al. (1995), Chen et al. (1995) and Chen (1995) in which optically thin solutions were found and studied. However, the main reason for the renewed interest in advection-dominated flows (e.g., Fabian & Rees 1995; Mineshige 1996) was the successful application of the model to the properties of the Galactic Center (Narayan et al. 1995) and the quiescent soft X-ray transients (Narayan, McClintock, & Yi 1996; see also Lasota 1996, Lasota, Narayan, & Yi 1996b). Recently Lasota et al. (1996a) applied advection-dominated flow models to the under-luminous active galactic nucleus NGC 4258.

Despite of their successful applications, models of optically thin accretion flows need to be improved, at least in order to test the validity of the simplifying assumptions that are often made. First, the description of the microphysics involved requires improvements and completion. For example, pair creation has been neglected until now, but it has been recently included into the model (Björnsson et al. 1996; Kusunose & Mineshige 1996). Second, the radial flow structure has been described in a simplified way. On the one hand Abramowicz et al. (1995) and Chen (1995), assumed that the flow is Keplerian (i.e. radially subsonic), but on the other hand, Narayan and collaborators use a self-similar solution which is significantly sub-Keplerian and subsonic everywhere. As shown by Chen et al. (1995) both type of models give the same general structure of the solutions but they differ in details. However, in the case of accretion on to a black hole the flow must be transonic (e.g., Abramowicz et al. 1988) so that neither of the descriptions is globally correct.

The self-similar solution may provide an excellent description of an advection-dominated flow far from the flow boundaries as it was shown by Narayan & Yi (1994) in the case of a settling subsonic solution (i.e. accretion on to a non-magnetized star). However, the structure of the flow in the inner and outer regions may have an important influence on the general properties of the modeled objects: high energy radiation is emitted primarily close to the black hole whereas optical and UV radiation will originate mostly in the outer regions where the flow may form a Keplerian disk.

In this paper we solve the problem of a transonic, advection-dominated flow on to a black hole and compare the solutions obtained with the local Keplerian and the self-similar solutions. The paper is organized as follows. In Section 2 we present the formulation of the problem. In Section 3 we obtain a self-similar solution of the Narayan & Yi (1994) type which will be compared with the general transonic solution. Section 4 contains the results of the numerical calculations. Finally in Section 5 we discuss the differences between our solutions and the self-similar and Keplerian approximations. Section 6 contains the conclusions.

## 2. FORMULATION

We consider an accretion disk which is axisymmetric, non-self-gravitating, optically thin and geometrically slim ($H/r \lesssim 1$) so that it can be described by vertically integrated equations. Here $r$ and $H$ are respectively the disk radius and the half-thickness. We will assume that the Keplerian rotational angular velocity is $\Omega_K = \sqrt{GM/r(r - r_g)^2}$ implied by the pseudo-Newtonian potential $\Phi = -GM/(r - r_g)$ (Paczyński & Wiita 1980), where $M$ is the mass of the central object and $r_g = 2GM/c^2$ is the Schwarzschild radius. Equations describing a stationary accretion flow are summarized below. Specifically, the continuity equation is

$$\dot{M} = -2\pi r v_r \Sigma, \tag{2.1}$$

where $\dot{M}$, $v_r$, and $\Sigma = 2H\rho$ are the mass accretion rate, the radial velocity and the surface density at a cylindrical radius respectively. Here $\rho$ is the mid-plane density of the disk. The equation of motion in the radial direction is

$$-v_r \frac{dv_r}{dr} + (\Omega^2 - \Omega_K^2)r - \frac{1}{\Sigma}\frac{dP}{dr} = 0, \tag{2.2}$$

where $\Omega$ is the angular velocity and $P$ is the vertically integrated pressure written, in a form compatible with the optically thin assumption, as

$$P = 2Hp = \frac{\Sigma \mathcal{R} T}{\mu}, \tag{2.3}$$

where $T$ and $p$ are the mid-plane temperature and pressure respectively, $\mu$ is the mean molecular weight, which for cosmic abundances is equal to 0.62, and $\mathcal{R}$ is the gas constant. The angular momentum equation can be integrated once with respect to the radius to yield the relation

$$r^2 \nu \frac{d\Omega}{dr} = v_r(\ell - \ell_{in}), \tag{2.4}$$

where $\nu$ is the kinematic viscosity and $\ell = \Omega r^2$ is the specific angular momentum. Here $\ell_{in}$ is the specific angular momentum lost from the disk, so that the viscous torque vanishes at the inner edge of the disk. The energy conservation equation is expressed by the balance between the local viscous heating, $Q_+$, the local radiative cooling, $Q_-$, and the global heat transport (radial advection), $Q_{\text{adv}}$. It is expressed as

$$Q_+ = Q_- + Q_{\text{adv}}. \tag{2.5}$$

The viscous heating rate per unit area is given by

$$Q_+ = \nu\Sigma\left(r\frac{d\Omega}{dr}\right)^2. \tag{2.6}$$

The advection cooling rate is taken in a form (e.g., Chen & Taam 1993):

$$Q_{\text{adv}} = \frac{\Sigma v_r}{r}\frac{P}{\Sigma}\left[\frac{4 - 3\beta}{\Gamma_3 - 1}\frac{d\ln T}{d\ln r} - (4 - 3\beta)\frac{d\ln\Sigma}{d\ln r}\right] \equiv \frac{\dot{M}}{2\pi r^2}\frac{P}{\Sigma}\xi, \tag{2.7}$$

where $\beta = P_g/P$ is the ratio of the gas pressure to the total pressure, $\Gamma_3 = 1 + (4 - 3\beta)(\gamma - 1)/[\beta + 12(\gamma - 1)(\beta - 1)]$, and $\gamma$ is the ratio of specific heats. We will take $\beta = 1$ and $\Gamma_3 = \gamma$. One should note that, due to different approaches to the vertical integration, the resulted $Q_{adv}$ is different from the one of Narayan & Yi (1994) as explained in Chen et al. (1995). The dimensionless advection factor $\xi$ characterises the entropy gradient. In our case, for a self-similar solution $\xi = 1$ (see more discussion later).

We shall assume that the local radiative cooling is provided by optically thin thermal bremsstrahlung with emissivity (erg s$^{-1}$ cm$^{-2}$),

$$Q_- = Q_{brem} = 1.24 \times 10^{21} H\rho^2 T^{1/2}. \tag{2.8}$$

Finally, we use a standard $\alpha$-model viscosity prescription (Shakura & Sunyaev 1973):

$$\nu = \frac{2}{3}\alpha c_s H, \tag{2.9}$$

where $\alpha$ is a constant, $c_s = \sqrt{P/\Sigma}$ is the local sound speed, and $H = c_s/\Omega_K$ is the local scale height which is also the half-thickness of the disk.

The above equations can be combined to reveal the presence of a sonic point. In particular, the equation of motion in the radial direction can be cast in the form

$$\frac{r}{v_r}\frac{dv_r}{dr} = \frac{N}{D}, \tag{2.10}$$

where

$$N = (\Omega_K^2 - \Omega^2)r^2 - C_s^2 + \frac{(\gamma - 1)r}{\Sigma v_r}\left(Q^+ - Q^-\right), \tag{2.11}$$

and

$$D = C_s^2 - v_r^2. \tag{2.12}$$

Here $C_s$ is the adiabatic sound speed ($c_s$ is the isothermal sound speed) defined as

$$C_s^2 = \gamma\frac{P}{\Sigma}. \tag{2.13}$$

The Mach number is defined as:

$$\mathcal{M} = |v_r|/C_s. \tag{2.14}$$

The vanishing of both $N$ and $D$ at the sonic point provides the regularity conditions required for a transonic solution of the flow structure. Furthermore, the differential equation for the temperature can be expressed as

$$\frac{r}{T}\frac{dT}{dr} = (1 - \gamma)\left(\frac{N}{D} - 1 - \frac{Q^+ - Q^-}{Pv_r}\right). \tag{2.15}$$

In the steady state accretion disk equations, the specific angular momentum, $\ell_{in}$, is an eigenvalue of the problem which is adjusted in such a way that a transonic solution is found at the sonic point. The equations are solved numerically using a relaxation technique subject to the regularity condition at the sonic point (Chen & Taam 1993). The initial trial solution is obtained from the integration of the equations (2.10) and (2.15) by assuming $\mathrm{d}\ln\Omega/\mathrm{d}\ln r = \mathrm{d}\ln\Omega_K/\mathrm{d}\ln r$ (see Paczyński & Bisnovatyi-Kogan 1981). The initial starting point for the integration is located away from the outer boundary used in the relaxation procedure. The exact location of the starting point does not affect the solution inside the outer boundary as long as it is sufficiently far away. Two kinds of advection-dominated solutions are used at the starting point. One is the Keplerian, advection-dominated solution of Abramowicz et al. (1995) and the other one is the self-similar solution (Narayan & Yi 1994). The trial solution of $v_r$ and $T$ at the outer boundary is then used as the boundary value in the relaxation procedure (with $\mathrm{d}\ln\Omega/\mathrm{d}\ln r \neq \mathrm{d}\ln\Omega_K/\mathrm{d}\ln r$). Typically, the method allows to converge to a solution after a few iterations.

At the sonic point, $v_r$ and T are calculated by requiring that both the numerator and the denominator in equation (2.10) vanish. Their derivatives are calculated by applying L'Hospital's rule. Defining $y = \mathrm{d}\ln|v_r|/\mathrm{d}\ln r$ at the sonic point, a quadratic equation for $y$ can be derived using equation (2.10) and (2.15). For example, in a polytropic approximation, it reads:

$$(\gamma + 1)y^2 + 2(\gamma - 1)y + 2a + \gamma - 1 = 0, \tag{2.16}$$

where,

$$a = \frac{r}{2C_s^2}\frac{\mathrm{d}(\Omega_K^2 - \Omega^2)}{\mathrm{d}r}. \tag{2.17}$$

(Note the difference with Chen & Taam (1993) where the last term $\gamma - 1$ is missing). If $y$ has no real solution, the sonic point is classified as a spiral point, which is not physical and is discarded. If the two solutions have different sign, the point is of saddle type, and if they have the same sign, the point is of nodal type (see Chen & Taam 1993 and references therein).

For definiteness, we assume that the black hole has a mass of $10\mathrm{M}_\odot$ and the inner numerical boundary of the disk is chosen to be $1.5\,r_g$. The outer numerical boundary of the disk is chosen to be typically about $1000\,r_g$, but larger in models in which we discuss the effects of boundary conditions. The results do not depend on the spatial resolution which is typically $\Delta r/r \sim 0.01$.

## 3. ANALYTIC SOLUTIONS

In some cases the differential accretion disk equations can be simplified to an algebraic system. For example, in the limit of Keplerian disk models (where all the pressure-gradient and inertial terms are neglected), one obtains the Shakura & Sunyaev equations and the corresponding analytic solutions. Narayan & Yi (1994) obtained self-similar solutions describing advection-dominated optically thin accretion flows. For the convenience of describing the numerical results in the next section, we derive the self-similar solution with the formulae outlined in §2. It should be mentioned that since the pseudo-Newtonian potential is used here, a self-similar solution is adequate only for large radii where the Keplerian rotation law can be approximated as $\Omega_K = \sqrt{GM/r^3}$. First we assume that each physical quantity $X$ can be expressed as (Narayan & Yi 1994),

$$X = Cr^N, \tag{3.1}$$

where $C \equiv C(\dot{M}, \alpha)$. The power index $N$, which may be different for each physical quantity, is constant. For the advection-dominated optically thin accretion flows considered here, the power indices for $v_r$, $\Omega$, and $T$ are $-1/2$, $-3/2$, and $-1/2$ respectively. The solutions are:

$$v_r = -\frac{5 + 2\varepsilon'}{2\alpha} g\Omega_K r, \tag{3.2}$$

$$\Omega = \sqrt{\frac{5 + 2\varepsilon'}{2\alpha^2}} g\Omega_K, \tag{3.3}$$

$$T = \frac{\mu}{\mathcal{R}} \frac{5 + 2\varepsilon'}{2\alpha^2} g(\Omega_K r)^2, \tag{3.4}$$

$$\mathcal{M} = \sqrt{(5/2 + \varepsilon')g/\gamma}, \tag{3.5}$$

where,

$$f\varepsilon' = \varepsilon = \frac{1 - \gamma/3}{\gamma - 1}, \tag{3.6}$$

$$g = \sqrt{1 + \frac{8\alpha^2}{(5 + 2\varepsilon')^2}} - 1. \tag{3.7}$$

Here, $f$ is the fraction of the advective cooling, and we have assumed $f = 1$ for the self-similar solutions. Note that $\varepsilon$ is different from that of Narayan & Yi (1994). This is due to a different definition of the advection term that was mentioned above. Note that, in the self-similar solutions, the dimensionless advection factor $\xi$ is determined by $\gamma$ only: $\xi = 1/(\gamma - 1) - 1/2$. Therefore, $\xi = 1$ for $\gamma = 5/3$ independent on the value of $f$ which one may have assumed.

All other quantities such as $p$ and $H$ can then be calculated accordingly. The self-similar solution has its limitations since no boundary condition has been taken into account. It has a constant Mach number and therefore the transonic region can not be described. However, as it will be shown later, it describes correctly the true solution asymptotically at large radii if the flow is advection-dominated.

## 4.  GLOBAL NUMERICAL SOLUTIONS

We have solved the full set of accretion disk differential equations numerically, using the method described in the previous section. The calculated models are summarized in Table 1 in terms of the two dimensionless input parameters: mass accretion rate $\dot{m} = \dot{M}/\dot{M}_E$, and the viscosity parameter $\alpha$. Here $\dot{M}_E = 4\pi GM/(c\kappa_{es})$ is defined as the Eddington accretion rate with $\kappa_{es} = 0.34$. Table 1 gives the eigenvalue $\ell_{in}$ [in the unit of $\ell_K(3\,r_g)$] of the solution, the location of the sonic point $r_s$, and the topological type of the sonic point.

The form of the solution depends on the values of $\alpha$ and $\dot{m}$. Abramowicz et al. (1995), Narayan & Yi (1995ab) and Chen et al. (1995) found that for viscosity parameters $\alpha \lesssim \alpha_{\rm crit}$, where $\alpha_{\rm crit}$ is a critical value depending on $r$ and on the assumed microphysics (see Chen et al. 1995, Björnsson et al. 1996), there exist a maximum accretion rate above which no optically thin solution exists. In the case of bremsstrahlung radiative cooling one gets (Abramowicz et al. 1995):

$$\dot{m}_{\rm max} \approx 1.7 \times 10^3 (r/r_g)^{-1/2}\alpha^2\xi^{-2} \tag{4.1}$$

For very small $\alpha$, since $\dot{m}_{\rm max} \propto \alpha^2 r^{-1/2}$, the optically thin solutions cease to exist for some $r > r_{\rm out}$, unless $\dot{m}$ is very low. For this reason we could find global solutions only for not very small $\alpha$ or for very low accretion rates.

We show in Figure 1 the results for a fixed value of $\alpha = 0.1$ and two accretion rates: $\dot{m} = 10^{-5}$ (Model 1, heavy solid line) and $\dot{m} = 10^{-2}$ (Model 2, heavy dashed line). The values of accretion rates are well below $\dot{m}_{\rm max}$ so that both solutions represent typical optically thin advection-dominated accretion flows (ADAFs). The form of the specific angular momentum is similar to that of the optically thick transonic accretion disks (see Abramowicz et al. 1988; Chen & Taam 1993). The solution is everywhere sub-Keplerian because of the high value of $\alpha$ (see Abramowicz et al. 1988). The sub-Keplerian character of the flow is also seen in the absence of a pressure maximum (note that the quantity plotted is the pressure $p$, not the vertically integrated pressure $P$ which always reach a maximum point independent on $\alpha$). $\xi$ is everywhere close to 1 and the self-similar solution is a good approximation for $r \gtrsim 30\,r_g$. The $\dot{m} = 10^{-2}$ solution is not as advectively dominated as the $\dot{m} = 10^{-5}$ one but it is still a robust ADAF. $\xi$ is practically the same for both accretion rates confirming the analysis of Chen (1995).

The dependence of the properties of the solution on $\alpha$ is illustrated in Figure 2. Here, the accretion rate is fixed as $\dot{m} = 10^{-5}$ and we consider two values of the viscosity parameter, $\alpha = 0.01$ (Model 3) and $\alpha = 0.001$ (Model 4). For such low values of $\alpha$ the specific angular momentum becomes super-Keplerian in the inner disk regions. Accordingly a pressure maximum is present. For the higher value of $\alpha$ the flow is advection-dominated and $\xi$ differs from 1 only in the innermost parts of the flow. The character of the inner flow for both $\alpha = 0.01$ and $\alpha = 0.001$ is different from that of $\alpha = 0.1$. In the first case (small $\alpha$), the angular momentum $\ell$ in the transonic region becomes almost constant with radius so that the viscous torque is very weak. This is therefore a 'relativistic Roche-lobe overflow' type, pressure driven, accretion (see Abramowicz et al. 1988 for a detailed explanation). In the second case, there is a gradient in $\ell$ and so a viscosity is needed for the mass to be accreted. This type of accretion could be therefore called 'viscosity dominated'.

One should also note that the sonic radius moves *inwards* with decreasing $\alpha$.

For $\alpha = 0.001$ the situation is still different. Only the form of the angular momentum as a function of $r$ is similar. The solution however ceases to be advection-dominated at $r \approx 30\,r_g$ and $\xi$ is nowhere well represented by $\xi = 1$. This is because, for this small $\alpha$, the accretion rate $\dot{m} = 10^{-5}$ is near the maximum value for these radii, so the local cooling becomes important. The self-similar approximation is still a good representation of $p$ for $r \gtrsim 30\,r_g$ but since $\xi$ is nowhere constant a self-similar solution is not a globally valid representation of the flow. It is seen that the local cooling tends to reduce $\xi$. In fact, for a radiative cooling dominated optically thin Keplerian disk, $\xi$ is negative (Wandel & Liang 1991).

Finally, we present a case (see Fig. 3, Model 5) in which the sonic point is of nodal type (in all previous solutions the sonic point was of saddle type). In this case $\alpha = 1$ and $\dot{m} = 10^{-3}$. Because of the large value of $\alpha$ the flow is very sub-Keplerian and the sonic point is far away from the black-hole. $\xi$ is always bigger than 1 but all the particularity of this solution are due to the high viscosity, the type of the sonic point has no obvious manifestations.

## 5. DISCUSSION

In the inner transonic region the optically thin ADAFs have properties similar to those of the optically thick, radiation-pressure dominated ADAFs (e.g., Abramowicz et al. 1988; Chen & Taam 1993). The character of the flow in this region is mainly determined by the Euler equation (Eq. [2.2]). This equation can be approximately rewritten as:

$$\frac{1}{\Sigma}\frac{\mathrm{d}P}{\mathrm{d}r} \approx \Omega_K^2 r\left[\left(\frac{\Omega}{\Omega_K}\right)^2 - 1 + \mathcal{M}^2\gamma\left(\frac{H}{r}\right)^2\right]. \tag{5.1}$$

For $r \gg r_s$, the right-hand-side of this equation is always negative since $\Omega < \Omega_K$ and the third term is negligible. If the flow becomes super-Keplerian (in the case of small $\alpha$) the rhs will become positive so that $P$ has a maximum. Even if the flow remains sub-Keplerian (in the case of large $\alpha$), our calculations show that the rhs will still become positive since the third term becomes important near $r = r_s$ where $(H/r)^2 \approx 1$ and $\mathcal{M} \gtrsim 1$. Therefore $P$ has always a maximum. The equation for pressure $p$ is

$$\frac{1}{\rho}\frac{\mathrm{d}p}{\mathrm{d}r} \approx \Omega_K^2 r\left[\left(\frac{\Omega}{\Omega_K}\right)^2 - 1 + (\mathcal{M}^2 - 1)\gamma\left(\frac{H}{r}\right)^2\right], \tag{5.2}$$

so that for globally sub-Keplerian flows $\mathrm{d}p/\mathrm{d}r$ is always negative and a pressure maximum is present only when the flow becomes somewhere super-Keplerian.

The topology of the solutions describing rotating accretion flows on to a black hole (Chen et al. 1995) implies a different behaviour of the optically thin and optically thick ADAFs at large radii.

In the optically thick case, the thermal equilibria of the disk at a given radius form a characteristic S-shaped curve on the $\Sigma - \dot{m}$ plane. Optically thick ADAFs (i.e., slim disks) correspond to the upper branch of the S-curve (see Abramowicz et al. 1988; Chen & Taam 1993). For a fixed accretion rate the S-curves move up (towards higher $\dot{m}$) with increasing radius (e.g., Fig. 2. of Chen & Taam 1993), so that at sufficiently large radii the $\dot{m} = const$ solution will correspond to an optically thick, gas-pressure dominated, Shakura-Sunyaev solution. Between the inner ADAF and the outer stable Keplerian solution there will be, however, an unstable region corresponding to the middle branch of the S-curve, so that such global solutions are thermally and viscously unstable.

In the optically thin case the solution structure is different. Contrary to the optically thick case where each value of the accretion rate corresponds to exactly one solution, the optically thin disk has *two* solutions for *one* given $\dot{m} < \dot{m}_{\max}$: an optically thin ADAF and a locally cooled optically thin solution.

Strictly speaking, the self-similar solution can never be an asympmtotic solution for an ADAF (at least in the case of bremsstrahlung cooling) because in a self-similar solution $Q_+ = Q_{\mathrm{adv}} \propto r^{-3}$ while $Q_- \propto r^{-5/2}$. For large values of $\alpha$ the relevant radius at which local cooling becomes important is too large to be of interest but for small viscosity parameters this radius could be $r < 10^3 - 10^4$ as it was shown in the previous section.

The dimensionless advection factor, $\xi$, is shown to be dependent on the fraction of the advective cooling, $f$, since for $f < 1$ the solution is no longer self-similar. In global solutions, the relation between $\xi$ and $f$ is nonlinear, but approximately, the smaller the $f$ the smaller the $\xi$. Furthermore, the value of $\xi$ calculated from the self-similar solution is also not valid in the transonic region even if $f \approx 1$. This is because a transonic solution is not self-similar.

In general, for large enough radii, the solution ceases to exist due to the accretion rate of the model exceeding the local maximum rate allowed for such kind of flows.

Solutions describing optically thin ADAFs can always be, formally, asymptotically Keplerian. In Figure 4 we show solutions in which the outer radius was fixed at either $r = 2500 \, r_g$ (heavy lines) or $r = 9900 \, r_g$ (thin lines), and the disk parameters are $\alpha = 0.01$ and $\dot{m} = 10^{-5}$ (Model 3). The solid lines represent the solution in which a self-similar solution (totally advection-dominated) is imposed at the outer boundary; and the dashed lines represent the solution in which a Keplerian local type advection-dominated solution is imposed at the outer boundary. Since ADAFs are asymptotically self-similar, the Keplerian boundary condition is rather far from the angular momentum corresponding to a self-similar flow. That is why the asymptotically Keplerian solution takes longer to relax than the asymptotically self-similar one which relaxes in a thin transition zone.

## 6. CONCLUSIONS

Our calculations of global transonic solutions describing ADAFs show that the self-similar solution is generally a good approximation of the full solution as long as the flow is fully advection-dominated and far away from the transonic region. Only when the accretion rate is close to the maximum one, the solution, as expected, ceases to be advection-dominated and a simple self-similar solution is nowhere close to the full solution.

In some applications (e.g., Narayan et al. 1996; Lasota et al. 1996b) one requires the presence of a cold Keplerian accretion disk at some distance $r \gtrsim 10 - 10^4 \, r_g$. The transition radius between the two types of flows will be given by physical processes that lead to the formation of an ADAFs (Meyer & Meyer-Hofmeister 1994; Narayan et al. 1996) This problem was not considered in the present work but the results presented above are a step towards its solution.

Ramesh Narayan, Shoji Kato and Fumio Honma obtained independently global solutions for fully advection dominated accretion flows. We thank Andrew King for inspiring remarks.

## REFERENCES

??bramowicz, M. A., Chen, X., Kato, S., Lasota, J.-P., & Regev, O. 1995, ApJ, 438, L37

??bramowicz, M. A., Czerny, B., Lasota, J.-P., & Szuszkiewicz, E. 1988, ApJ, 332, 646

??bramowicz, M., Lasota, J.-P., & Xu, C. 1986, in IAU Symp. 119, Quasars, ed. G. Swarup & V. K. Kapahi (Dordrecht: Reidel), 376

??egelmann, M. C. 1978, MNRAS, 184, 53

??———. 1979, MNRAS, 187, 237

??egelman, M. C. & Meier, D. L. 1982, ApJ, 253, 873

??ertout, C., Basri, G., & Cabrit, S. 1991, in The sun in time, ed. C. P. Sonett, M. S. Giampapa, & M. S. Matthews (Tucson: University of Arizona Press), 682

??jörnsson, G., Abramowicz, M. A., Chen, X., & Lasota, J.-P. 1996, in preparation

??hen, X. 1995, MNRAS, 275, 641

??hen, X., Abramowicz, M. A., Lasota, J.-P., Narayan, R., & Yi, I. 1995, ApJ, 443, L61

??hen, X., & Taam, R. E. 1993, ApJ, 412, 254

??abian, A. C., & Rees, M. J. 1995, MNRAS, submitted

??rank, J., King, A. R., & Raine, D. 1992, Accretion Power in Astrophysics (Cambridge: CUP)

??usunose, M., & Mineshige, S. 1996, in preparation

??asota, J.-P. 1996, in IAU Colloquium 158, Cataclysmic Variables and Related Objects, ed. J. Wood & A. Evans (Dordrecht: Kluwer), in press

??asota, J.-P., Abramowicz, M. A., Chen, X., Krolik, J. H., Narayan, R., & Yi. I. 1996a ApJ, in press

??asota, J.-P., Narayan, R., & Yi. I. 1996b, in preparation

??eyer, F., & Meyer-Hofmeister, E. 1994, A&A, 288, 175

??ineshige, S. 1996, PASJ, submitted

??arayan, R., McClintock, J. E., & Yi, I. 1996, ApJ, , in press

??arayan, R., & Popham, R. 1993, Nature, 362, 820

??arayan, R., & Yi, I. 1994, ApJ, 428, L13

??——. 1995a, ApJ, 444, 231

??——. 1995b, ApJ, 452, 710

??arayan, R., Yi, I., & Mahadavan R. 1995, Nature 374, 623

??aczyński, B., & Bisnovatyi-Kogan, G. 1981, Acta Astr., 31, 283

??aczyński, B., & Wiita, P. J. 1980, A&A, 88, 23

??hakura, N. I., & Sunyaev, R. A. 1973, A&A, 24, 337

??andel, A., & Liang, E. P. 1991, ApJ, 380, 84

Fig. 1.— Disk solutions for Model 1 ($\alpha = 0.1$ and $\dot{m} = 10^{-5}$, solid lines) and Model 2 ($\alpha = 0.1$ and $\dot{m} = 10^{-2}$, dashed lines). The heavy lines are the global solution and the thin lines are the corresponding self-similar solution. Note that $\mathcal{M}$, $\Omega$, and $\xi$ depend on $\dot{m}$ very weakly.

Fig. 2.— Disk solutions for Model 3 ($\alpha = 0.01$ and $\dot{m} = 10^{-5}$, solid lines) and Model 4 ($\alpha = 0.001$ and $\dot{m} = 10^{-5}$, dashed lines). The heavy lines are the global solution and the thin lines are the corresponding self-similar solution. Note the super-Keplerian angular momentum and the maximum pressure near the transonic region. Note also that in Model 4, the local cooling becomes important for large radii.

Fig. 3.— Disk solutions for Model 4 ($\alpha = 1.0$ and $\dot{m} = 10^{-3}$). The heavy lines are the global solution and the thin lines are the corresponding self-similar solution. This solution is very sub-Keplerian and the sonic point is a nodal type point.

Fig. 4.— The outer boundary effects shown for Model 3 ($\alpha = 0.01$ and $\dot{m} = 10^{-5}$). The heavy and thin lines are for the outer radius fixed at $r = 2500\, r_g$ and $r = 9900\, r_g$ respectively. The solid lines represent the solution in which a self-similar solution (totally advection-dominated) is imposed at the outer boundary; and the dashed lines represent the solution in which a Keplerian local type advection-dominated solution is imposed at the outer boundary.

| Model | $\alpha$ | $\dot{M}/\dot{M}_{\mathrm{E}}$ | $\ell_{in}/\ell_K(3\,r_g)$ | $r_s/r_g$ | sonic point |
|-------|----------|-------------------|--------------------------|-----------|-------------|
| 1 ...... | 0.1 | $10^{-5}$ | 0.92967 | 2.47 | saddle |
| 2 ...... | 0.1 | $10^{-2}$ | 0.92983 | 2.47 | saddle |
| 3 ...... | 0.01 | $10^{-5}$ | 1.0500601 | 2.112 | saddle |
| 4 ...... | 0.001 | $10^{-5}$ | 1.0740025 | 2.048 | saddle |
| 5 ...... | 1.0 | $10^{-3}$ | 0.446 | 7.82 | nodal |

Table 1: Model Sequences